\documentclass{PoS}

\newcommand{\KLnu}{K_L \to \pi^0 \nu \bar \nu}
\newcommand{\Kp}{K^+ \to \pi^+ \nu \bar \nu}
\newcommand{\KLl}{K_L \to \pi^0 \ell^+ \ell^-}     
\newcommand{\KLe}{K_L \to \pi^0 e^+ e^-}
\newcommand{\KLmu}{K_L \to \pi^0 \mu^+ \mu^-}
   
\newcommand{\gev}{\,{\rm GeV}}   
\newcommand{\tev}{\,{\rm TeV}}   
\newcommand{\be}{\begin{equation}}   
\newcommand{\ee}{\end{equation}}   
\newcommand{\bea}{\begin{eqnarray}}   
\newcommand{\eea}{\end{eqnarray}}

\title{Rare Kaon Decays Beyond the Standard Model}

\ShortTitle{Rare Kaon Decays Beyond the SM}

\author{\speaker{Cecilia Tarantino}\thanks{It is a pleasure to thank the
        organizers of ``Kaon'07'' for the very interesting
        conference realized at the Laboratori Nazionali di Frascati. I am also
        grateful to Andrzej~J.~Buras and Federico Mescia for inspiring
        some important issues of my talk and to Ulrich Haisch for a careful
        reading of the manuscript. This work is partially supported by the Cluster of Excellence `Origin and Structure of the Universe' and by the German `Bundesministerium f\"ur Bildung und Forschung' under contract 05HT6WOA.}\\
        Physik Department, Technische Universit\"at M\"unchen, D-85748 Garching, Germany\\
        E-mail: \email{tarantino@fis.uniroma3.it}}

\abstract{The rare kaon decays $\KLnu$, $\Kp$, $\KLe$ and $\KLmu$ are
  theoretically very clean and, being strongly CKM suppressed, highly
  sensitive to New Physics (NP). Recent Flavour Physics analyses show that they
  represent unique probes for revealing NP effects and to provide information
  on the NP flavour structure.
After a brief discussion of the main properties that make rare $K$ decays
  so promising and of the basic ideas of the most interesting NP models, we
  review the results of recent phenomenological analyses both within and
  beyond the framework of Minimal Flavour Violation (MFV), where the sources of
  flavour violation are the same as in the Standard Model.
Within MFV we present the expectations found for rare $K$ decays from a
  model-independent analysis and in three MFV models: the Littlest
  Higgs (LH) model, the (extra-dimension) Appelquist-Cheng-Dobrescu model and the
  Minimal Supersymmetric Standard Model (MSSM) with MFV. Beyond MFV, we
  discuss the results recently
  found within the MSSM (without MFV), the LH model
  with T-parity (LHT) and the 3-3-1 ($Z'$) model.
While in MFV models only small ($<30$\%) NP effects are allowed in the branching ratios
  of $\KLnu$, $\Kp$, $\KLe$ and $\KLmu$, beyond MFV, in particular in the
  MSSM and in the LHT model, large (up to an order of magnitude)
  enhancements w.r.t. the SM turn out to be possible.}

\FullConference{KAON International Conference\\
                 May 21-25 2007\\
                 Laboratori Nazionali di Frascati dell'INFN, Rome, Italy}

\begin{document}

\section{Rare Kaon Decays as Unique Probes of New Physics}
\label{sec:1}
The rare kaon decays $\KLnu$, $\Kp$, $\KLe$ and $\KLmu$ play a fundamental
role in looking for New Physics (NP) beyond the Standard Model (SM), because
they are theoretically very clean and highly sensitive to NP, being strongly
suppressed within the SM.\footnote{For an extensive discussion of the
  SM theoretical predictions of rare kaon decays and for a complete list of
  references we address the reader to ref. \cite{SM}.}
Moreover, their theoretical cleanness is not due to any
cancellation and remains true also beyond the SM, thus allowing a precise
investigation of NP effects.

The decays $K \to \pi \nu \bar \nu$ are short-distance dominated and involve 
matrix elements that can be accurately extracted from the experimental data on
$K_{\ell 3}$ decays \cite{Marciano:1996wy,Mescia:2007kn}, such that the
intrinsic theoretical uncertainty is smaller than $5$\%.
In $\KLnu$, as $K_L \simeq (K^0 + \bar K^0)/\sqrt{2}$ and the $K^0 \to \pi^0$
and $\bar K^0 \to \pi^0$ matrix elements differ by a minus sign, the CKM
contributions involved are the CP-violating imaginary parts
${\rm Im}(\lambda_i)\equiv {\rm Im}(V_{id} V_{is}^*)$ ($i=u,c,t$). Through the GIM
mechanism ${\rm Im}(\lambda_u)$ can be expressed in terms of ${\rm Im}(\lambda_t)$ and
${\rm Im}(\lambda_c)$, both strongly Cabibbo suppressed (${\rm Im}(\lambda_t)
\sim {\rm Im}(\lambda_c) \sim \mathcal{O}(\lambda^5)$).
Moreover, the short-distance loop-function $X$ describing $\KLnu$ is highly
suppressed if a small quark mass runs in the loop, thus making the charm
contribution negligible.
It turns out, then,  that $\KLnu$ is CKM suppressed as
$\mathcal{O}(\lambda^5)$ and fully dominated by the top contribution.
The decay $\Kp$ is also described by the short-distance function $X$ but, in
the $K^+$ decay,  both imaginary and real parts of the CKM contributions are 
involved and, being ${\rm Re}(\lambda_c) \sim {O}(\lambda) \gg {\rm
  Re}(\lambda_t)  \sim {O}(\lambda^5)$, the 
charm contribution cannot be neglected in this case.
Its perturbative calculation requires some care as the charm scale ($m_c \sim
1 \gev$) is not as high as the top one ($m_t \sim 10^2 \gev$).
Recently, however, the uncertainty on the charm contribution has been significantly reduced thanks to the computation of the NNLO corrections \cite{Buras:2005gr}
and the study of long-distance and dimension-8
operator effects \cite{Isidori:2005xm}.
This improvement, together with the determination \cite{Mescia:2007kn} of the
operator matrix elements for both $K \to \pi \nu \bar \nu$ and $\KLl$ from
$K_{\ell 3}$ data using chiral perturbation
theory beyond LO, represents the main theoretical progress in rare $K$ decays.

The decays $\KLl$ ($\ell=e, \mu$), though not as theoretically clean as the
golden modes $K \to \pi \nu \bar \nu$, have an intrinsic theoretical uncertainty
smaller than $10$\% and represent very promising channels for their
peculiar sensitivity to NP.
They involve three contributions of comparable size: a
direct CP-violating contribution of short-distance origin, an indirect 
CP-violating contribution that can be determined from the experimental
measurement of $K_S \to \pi^0 \ell^+ \ell^-$ and a long-distance CP-conserving contribution that
can be determined from the experimental data on $K_L \to \pi^0 \gamma
\gamma$.
This last contribution comes from the two-photon channel $K_L \to \pi^0 (\gamma
\gamma)_{J=0,2} \to \pi^0 (\ell^+ \ell^-)_{J=0,2}$ that, being of
long-distance origin, represents the main source of uncertainty, now under
control thanks to recent studies \cite{Buchalla:2003sj,Isidori:2004rb}.
Moreover, the two-photon contribution distinguishes between
the muon and electron modes as the $J^{\rm CP}=0^{++}$
channel is helicity suppressed and, therefore, significant only in the
(heavier) muon case.
More in general, a relevant peculiarity of  the $\KLl$ decays is that the different impact of
helicity suppressed contributions to the muon and electron modes makes the
comparison of $\KLmu$ and $\KLe $ a powerful tool to
reveal the properties of NP \cite{Mescia:2006jd}.

Finally, we collect in table
\ref{tab:values} the SM theoretical
predictions for the four rare kaon decays in question compared to present
experimental measurements or limits, to underline
the accuracy of the SM predictions and
the large room for NP still allowed by the experiments.   
\begin{table}[!]
\begin{center}
\begin{tabular}{||c||c|c||} \hline
 Decay & SM & Experiment \\ \hline
$\Kp$ & $(7.8 \pm 0.8)\cdot 10^{-11}$ \cite{Mescia:2007kn} & $(14.7^{+13}_{-8.9})\cdot
10^{-11}$ \cite{expK+} \\ 
$\KLnu$ & $(2.5 \pm 0.4)\cdot 10^{-11}$ \cite{Mescia:2007kn} & $< 2.1\cdot
10^{-7}$ \cite{Ahn:2006uf} \\ 
$\KLe$ & $(3.5^{+1.0}_{-0.9})\cdot 10^{-11}$ \cite{Mescia:2006jd} & $< 2.8\cdot
10^{-10}$ \cite{Alavi-Harati:2003} \\
$\KLmu$ &  $(1.4 \pm 0.3)\cdot 10^{-11}$ \cite{Mescia:2006jd} & $< 3.8\cdot
10^{-10}$ \cite{Alavi-Harati:2000} \\ 
\hline
\end{tabular}
\end{center}
\vspace*{-0.3cm} 
\caption{\sl SM predictions and experimental measurements of the four rare kaon decays}
\vspace*{-0.2cm} 
\label{tab:values}
\end{table}

\section{The Most Interesting NP Models}
\label{sec:2}
\vspace*{-0.2cm}
\subsection{Distinction between Models with and without Minimal Flavour Violation}
\label{subsec:2.1}
In this section we discuss the basic features of the most appealing NP models,
i.e. the NP models that solve at least some of the open questions of the SM (first of all the so-called little hierarchy problem), remaining simple
enough to be predictive and compatible with electroweak (ew) precision tests and with flavour constraints.
We recall that the little hierarchy problem occurs as the NP
scale is required to be low ($\Lambda_{\rm NP} \sim 1 \tev$) to
explain the lightness of the Higgs mass without
fine-tuning, while it is constrained to be quite
high ($\Lambda_{\rm NP} \sim 5-10 \tev$) by phenomenological analyses where NP 
effects are taken into account by higher dimensional operators \cite{Barbieri:2000gf,Bona:2007vi}.

NP models can satisfy or not the Minimal Flavour
Violation (MFV) hypothesis, consisting in the absence of new sources of flavour
violation in addition to the SM.
MFV provides a natural, though not very 
exciting, explanation to the great success of 
the SM in Flavour Physics. 
There are two different formulations of MFV: a {\it pragmatic} and a {\it
theoretical} one.
The {\it pragmatic} formulation \cite{Buras:2000dm} states that a model is a
MFV model if it satisfies
two constraints: the only source of flavour violation is the CKM matrix (as in
the SM) and  it involves only the SM operators. Because of this second
property, the {\it pragmatic}
formulation is also called Constrained MFV (CFMV) \cite{BBGT}.
The {\it theoretical} formulation \cite{D'Ambrosio:2002ex} consists in building a MFV model by
using as flavour violating building blocks only the SM Yukawa couplings.
Both formulations are valid, similar indeed, with a slight
difference due to the absence of new operators in
the CMFV case.
Finally, NP models going beyond MFV are those that introduce new sources of
flavour violation in addition to the SM.

In order to see how NP effects can appear in Flavour Physics observables, let
us consider a generic weak amplitude $A$ that can be schematically
written as \cite{Buras:2001tc}
\be
A=V^i_{\rm CKM}\, \{B_i\,\eta^i_{\rm QCD}\,[F^i_{\rm SM} +
F^i_{\rm CMFV}]+B_i^{\rm MFV}\,(\eta^i_{\rm QCD})^{\rm MFV}
\,F^i_{\rm MFV}\}+V^i_{\rm NEW}\,B_i^{\rm NEW}\,(\eta^i_{\rm QCD})^{\rm NEW}
\,F^i_{\rm NEW}\,,
\label{eq:weakA}
\ee
where $F^i$ are the (perturbative) short-distance functions\footnote{The short-distance function involved in $\KLnu$ and $\Kp$ is the loop-function $X$
  calculated from Z-penguin and box diagrams. In $\KLl$, in addition to $X$, 
the loop-functions
  $Y$ and $Z$ are also involved, which require the calculation of the
  photon-penguin diagram.}, $\eta^i$ are their NLO corrections
in QCD, $B_i$ are the (non-perturbative) low energy parameters describing
the operator matrix elements and $V^i_{\rm CKM}$ denote the  CKM combinations
entering the decay.
Within CMFV the only NP effect is a modification of the short-distance
functions $F_i$, while in MFV new operators and therefore new
$B$-parameters can also appear. A more general NP model going beyond MFV, in addition, can
introduce new sources of flavour violation different from the CKM ones,  
denoted by $V^i_{\rm NEW}$.
\vspace*{-0.2cm}
\subsection{MFV Models}
\label{subsec:2.2}
Very interesting NP models that satisfy MFV are the Littlest Higgs (LH) model
\cite{Arkani-Hamed:2002qy}, the Appelquist-Cheng-Dobrescu (ACD) model
\cite{Appelquist:2000nn} and the Minimal
Supersymmetric Standard Model (MSSM) with MFV \cite{SUSY}.

The LH model \cite{Arkani-Hamed:2002qy} solves the little hierarchy problem, by interpreting the Higgs boson as a pseudo-Goldstone boson of a global
symmetry, that being massless at tree-level can be naturally kept light.
The global symmetry in question is $SU(5)$, spontaneously broken to $SO(5)$ at
a NP scale $f \sim \mathcal{O}(\tev)$.
The SM problematic quadratic corrections to the Higgs mass are cancelled in
the LH model by the contributions of new particles having the same
spin-statistics of the SM ones: heavy gauge bosons
($W_H^\pm,Z_H, A_H$), the heavy top ($T$) and a scalar
triplet ($\Phi$). 
These new particles, however, introduce tree-level corrections 
such that the NP scale is constrained to be quite large ($f > 2-3
\tev$) by ew precision tests and the model
becomes to be less appealing.

The ACD model \cite{Appelquist:2000nn} is characterized by one universal extra-dimension,
that means that the SM particles propagate in $5$ dimensions.
In $4$ dimensions the extra-dimension is taken into account by the appearance
of new (heavy) Kaluza-Klein (KK) modes. The model satisfies a discrete parity
called KK-parity that, implying an even number
of KK modes in vertices, avoids undesired tree-level corrections.
A very nice feature of this model is its simplicity. Typically, in fact,
flavour observables can be expressed at LO in terms of only one parameter: the compactification radius $R$.

The MSSM \cite{SUSY} is the supersymmetric extension of the SM with minimal matter content.
The new particles are the so-called sparticles, i.e. superpartners of the SM
ones with opposite spin-statistics and heavy masses (that cancel the
SM quadratic corrections to Higgs mass), and two Higgs doublets $H_u$ and
$H_d$ whose vacuum expectation values define the model parameter $\tan \beta
\equiv v_u/v_d$.
The R-parity, i.e. the discrete symmetry inducing an even number of sparticles in vertices, is introduced
in order to avoid undesired tree-level contributions, e.g. to the proton decay.
The MSSM can satisfy MFV once the soft breaking terms are imposed to be
adequate combinations of Yukawa couplings. 
Moreover, for small values of $\tan \beta$ the effects of new operators are
unimportant and the model satisfies CMFV (unless the MSSM parameter $\mu$
is large, $\mu > 500$ \cite{Altmannshofer:2007cs}).
At large $\tan \beta$, instead, the impact of new operators generated by Higgs
contributions becomes important.
\vspace*{-0.2cm}
\subsection{Non-MFV Models}
\label{subsec:2.3}
A very interesting NP model going beyond MFV is certainly the MSSM \cite{SUSY}
where the
soft breaking terms are left free to introduce new sources of flavour
violation. In particular, $27$ new flavour changing couplings appear in the
squark propagators with significant effects in flavour observables.

An interesting alternative to the MSSM is the LH model when a discrete
symmetry, T-parity \cite{tparity}, is introduced.
T-parity acts similarly to R-parity in supersymmetry or KK-parity in
extra-dimension models, assigning opposite parities to SM and new particles
and, thus, forbidding dangerous tree-level contributions.
The LH model with T-parity (LHT) satisfies ew precision tests already at a
quite low NP scale ($f > 500 \gev$), as required to solve the little hierarchy problem.
Once T-parity is introduced, new particles appear in addition to the heavy
gauge bosons, the heavy top and the scalar triplet already present without
T-parity.
They are the so-called mirror fermions, i.e. (same spin-statistics)
partners of the SM fermions, characterized by new flavour interactions with
important effects in flavour observables.

Another quite simple and predictive extension of the SM is represented by the
3-3-1 model \cite{331}, whose name comes from its extended gauge group $SU(3)_c
\times SU(3)_L \times U(1)_X$. The extension of the left-handed gauge group to
$SU(3)_L$ requires the introduction of new particles: heavy gauge bosons
and heavy fermions to complete fermion multiplets. Different fermion
generations can belong to different representations of $SU(3)_L$ and, indeed, 
the first two quark generations are set in
triplets, while the third quark generation in an
antitriplet. This choice is welcome as it explains the necessity of three
fermion generations in order to obtain anomaly cancellation and QCD asymptotic freedom. 
A consequence of the peculiarity of the third generation is that the neutral heavy
gauge boson $Z'$ (the model is also called $Z'$-model) can transmit flavour
changing neutral currents at tree-level, with important effects in flavour 
observables.

\section{Expectations for Rare Kaon Decays Beyond the SM}
\label{sec:3}
\vspace*{-0.2cm}
\subsection{Within MFV}
\label{subsec:3.1}
A model independent study of Flavour Physics within CMFV has been performed in
\cite{Bobeth:2005ck}, taking into account the available information from the Unitarity Triangle (UT)
analysis and from the measurements of $Br(B \to X_s \gamma)$ and  $Br(B \to
X_s \ell^+ \ell^-)$.
The goal of that study was the determination of the range allowed for the 
 short-distance function $C$ (i.e. the Z-penguin contribution, while NP
 effects in box and gluon-penguin diagrams, typically small in MFV models,
 were neglected) and the evaluation of upper bounds for rare decays.
It was found that the enhancements can be at most of $30$\% for $Br(\KLnu)$ and $Br(\Kp)$, 
whose accurate experimental measurements are
certainly looked forward in view of their high sensitivity to NP.
The analysis of \cite{Bobeth:2005ck} has just been updated and improved in \cite{UliAndi} by
including the constraints from the observables related to the $Z \bar b b$
vertex. 
The main effect of these new constraints is the exclusion of the negative
(non-SM) sign solution for the $C$ function.
As a consequence, $Br(\KLnu)$ and $Br(\Kp)$ turn out to be bounded also from
below and to lie in the following $95$\% C.L. ranges  \cite{UliAndi} 
\be
Br(\Kp)_{\rm CMFV} \in   [4.3, 10.7] \cdot 10^{-11}\,,\qquad
Br(\KLnu)_{\rm CMFV}  \in  [1.6, 4.4] \cdot 10^{-11}\,.
\label{eq:UliAndi}
\ee
Furthermore, in \cite{Bobeth:2005ck} a strong correlation between
$Br(\KLnu)$ and $Br(\Kp)$ has been pointed out. This is expected from the 
MFV relation \cite{D'Ambrosio:2002ex,Buras:2001af}
\be
\frac{Br(\KLnu)}{Br(\Kp)}\Big\vert_{\rm MFV} = 
  \frac{Br(\KLnu)}{Br(\Kp)}\Big\vert_{\rm SM} 
\cdot[1 + \epsilon_c \, {\rm sign}(X)]\,,
\label{eq:corr}
\ee
where two branches are distinguished by the sign of the 
 $X$ function and $\epsilon_c$ denotes a small charm contribution coming from
$\Kp$.
Within CMFV, the exclusion of the negative sign solution for $C$
\cite{UliAndi}\footnote{With the assumption of \cite{Bobeth:2005ck,UliAndi}
  that no NP effects appear in box diagrams,
  the NP contributions to the $X$ function coincide with those to the
  (Z-penguin) $C$ function and, therefore, ${\rm sign}(X) \equiv {\rm sign}(C)$.} leaves
only one branch in the correlation (\ref{eq:corr}).   

For completeness we mention the results found from the analyses of rare kaon
decays in specific CMFV models.
In the LH model, where the NP scale is constrained to be quite large ($f > 2-3
\tev$) by ew precision tests, NP effects in flavour observables turn
out to be suppressed and, in particular, in $Br(\KLnu)$ and $Br(\Kp)$ can
amount at most to $15$\% \cite{Buras:2006wk}.
Similar enhancements turn out to be allowed in the ACD model for
values of the compactification radius  $1/R \simeq 200 \gev$ \cite{Buras:2002ej}. Even
smaller enhancements ($<5$\%) are found \cite{Haisch:2007vb} if $1/R > 500-600 \gev$, as
required by the $Br(B \to X_s
\gamma)$ constraint after including the 
recent NNLO corrections \cite{Misiak:2006zs}.
In the MSSM with MFV and small ($<30$) $\tan \beta$, the NP effects
to $Br(\KLnu)$ and $Br(\Kp)$ turn out to be naturally small ($\sim 10$\%) \cite{Isidori:2006qy}, but
could saturate the model-independent bounds of \cite{Bobeth:2005ck,UliAndi}.
At large $\tan \beta$, charged Higgs loops generate a new  operator
$\bar s \gamma^\mu (1+\gamma_5) d \bar \nu \gamma^\mu (1-\gamma_5) \nu$, not
suppressed by any light quark mass. 
Nevertheless, it has been pointed out \cite{Isidori:2006jh} that within MFV its
contribution doesn't modify significantly the small $\tan
\beta$ results just mentioned above.

Concerning the decays $\KLl$, in models without new operators the
NP effects in  $\KLl$ are generally smaller than in $K \to \pi \nu \bar \nu$.
Within CMFV the enhancements of $Br(\KLl)$ are expected to be utmost of $10$\%.
Still, a combined measurement of electron and muon modes would be very
interesting to reveal helicity suppressed NP effects.

Finally, we emphasize that the absence, within CMFV, of large 
departures from the SM in rare kaon decays, though not very exciting, provides a strong message: if a
large enhancement was found in rare kaon decays then all
CMFV models would become very unlikely.   
\vspace*{-0.2cm}
\subsection{Beyond MFV}
\label{subsec:3.2}
We discuss now the expectations for rare kaon decays beyond MFV in the three NP
models briefly described in Section \ref{subsec:2.3}: the MSSM \cite{SUSY} with a
general flavour structure, the LHT model \cite{tparity} and the 3-3-1 ($Z'$)
model \cite{331}.

Various analyses of rare kaon decays have been performed \cite{Isidori:2006qy}-\cite{Buras:1999da} in the MSSM
with a general flavour structure.\footnote{For an extensive discussion and
  bibliography on
  rare kaon decays in the MSSM we address the reader to ref. \cite{smith}.}
In order to obtain simple analytic expressions for $s-d$ transitions,
involved in rare K decays, it is useful to work in a basis where the gauge
interactions of the SM down quarks with gauginos and neutralinos
are flavour diagonal. In this basis the squark mass matrices, made up of four 
blocks corresponding to the squark chiralities $LL$, $LR$, $RL$ and $RR$,
are non-diagonal in flavour space.
It is convenient to use the so-called Mass Insertion Approximation (MIA) to expand them in the
small off-diagonal entries $\delta$'s.
In rare kaon decays the dominant NP effects are represented by chargino
up-squark loops, with double insertions of $\delta^U_{LR}$ not suppressed by
$\mathcal{O}(M_W/M_{SUSY})$ nor CKM, and by charged-Higgs top-quark loops, with
$\delta^D_{RR}$ contributions not suppressed by any light quark mass and
enhanced by $\tan^4 \beta$.   

\begin{figure}[t]
\begin{center}
\includegraphics[scale=0.25,angle=270]{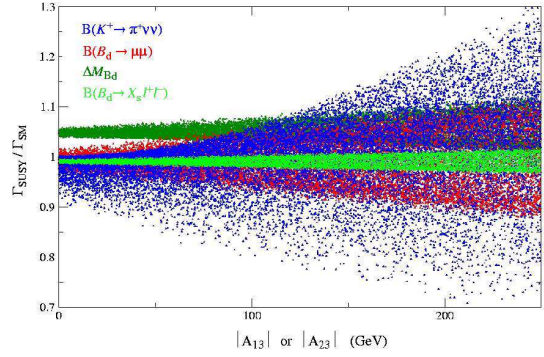}
\end{center}
\vspace{-0.5cm}
\caption{\small\sl Dependence, within the MSSM, of flavour observables on the trilinear couplings $A_{i3}$
  ($i=1,2$) or, equivalently, on the $\delta^U_{LR}$ mass insertions as with a good
  approximation $(\delta^U_{LR})_{i3} \propto m_t A_{i3}$ ($i=1,2$) \cite{Isidori:2006qy}.}
\vspace*{-0.2cm} 
\label{fig:MSSM}
\end{figure}

In \cite{Isidori:2006qy} it has been found that the mass insertions $\delta^U_{LR}$ are
essentially not constrained by present measurements of $B$ and $K$
observables, as shown in fig. \ref{fig:MSSM}.
From fig. \ref{fig:MSSM} it is also evident that rare kaon decays, e.g.
$Br(\Kp)$, are potentially the most constraining for the $\delta^U_{LR}$'s.
In \cite{Isidori:2006jh}, then, it has been found that precise measurements of $Br(\KLnu)$ and $Br(\Kp)$ could
also provide the most stringent bounds on the mass insertions $\delta^D_{RR}$.
Furthermore, the analysis of \cite{Isidori:2006qy}, where $\tan
\beta <30$, finds that $\mathcal{O}(1)$
enhancements of $Br(\KLnu)$ and $Br(\Kp)$ are possible.
From a very general scan \cite{Buras:2004qb} over the MSSM parameters, even enhancements
of one order of magnitude turn out to be possible. 
Concerning the decays $\KLl$, at large $\tan \beta$ the contributions of the
(pseudo)scalar operators $\bar s   d \bar \ell  \ell$ and $\bar s
 d \bar \ell \gamma_5 \ell$ are enhanced. As they are helicity suppressed,
 they affect the muon mode only and could provide a clear signal.

\begin{figure}[t]
\begin{center}
\includegraphics[scale=0.5]{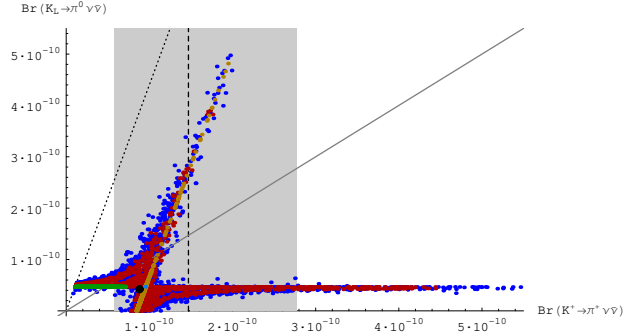}
\end{center}
\vspace{-1.0cm}
\caption{\it $Br(\KLnu)$ as a function of $Br(\Kp)$ in the LHT model \cite{Blanke:2006eb}. The shaded
    area represents the experimental $1\sigma$-range for $Br(\Kp)$. The
    model-independent Grossman-Nir bound \cite{Grossman:1997sk} is displayed by the dotted line, while the solid line
    separates the two areas where $Br(\KLnu) > (<) Br(\Kp)$.}
\vspace*{-0.2cm} 
\label{fig:LHT}
\end{figure}

We now discuss the results found in 
ref. \cite{Blanke:2006eb} for rare kaon decays within the LHT model.
As discussed in Section \ref{subsec:2.3}, after introducing 
T-parity the model allows for new sources of flavour violation induced by
mirror fermions. 
In particular, a new mixing matrix $V_{Hd}$ that
governs the interactions of a SM quark with a mirror quark and a heavy gauge
boson is involved in the quark sector.
It is parameterized in terms of three new mixing angles and three new
(CP-violating) phases \cite{Blanke:2006xr} and its hierarchy can be completely
different from the CKM.
As a consequence, the strong CKM suppression of rare kaon decays present in 
the SM can have no analogy in the mirror sector and large NP effects may occur.
Indeed, this is the main result found in \cite{Blanke:2006eb} and shown in
fig. \ref{fig:LHT}, where two branches in the $Br(\Kp)$-$Br(\KLnu)$ plane
result from a general scan over the LHT parameters.
They are selected by the indirect CP-violation parameter $\varepsilon_K$,
which turns out to be the most stringent constraint.
The {\it horizontal} branch corresponds to an enhancement of  $Br(\Kp)$ up to a
factor $5$, with $Br(\KLnu)$ close to the SM prediction.
The {\it oblique} branch, instead, signals that an enhancement of $Br(\KLnu)$
up to a factor $10$ is possible, with a simultaneous enhancement of $Br(\Kp)$
of at most a factor $3$.
In \cite{Blanke:2006eb} the decays $\KLl$ have also been studied, finding
enhancements up to a factor $2$ w.r.t. the SM and a strong
correlation between the muon and the electron modes and with $\KLnu$. 
Another important result of ref. \cite{Blanke:2006eb} concerns the UT angle $\beta$, for
which a difference between the value obtained from
$B_d \to J_\psi K_s$ and a future determination from $\KLnu$ 
would be a clear signal of NP.
Within the LHT model the universality between $B$ and $K$ systems is
 strongly violated and such a signal could be easily seen.
More recently, the potentiality of the direct
CP-violating parameter $\varepsilon'/\varepsilon$ in constraining rare kaon
decays within the LHT model has been studied \cite{Blanke:2007wr}. 
While the experimental measurement of $\varepsilon'/\varepsilon$ is accurate
at the $10$\% level ($(\varepsilon'/\varepsilon)_{\rm exp.}=(17 \pm 2)
\cdot 10^{-4}$  \cite{epsexp}), the theoretical prediction requires to know
the value for the gluon penguin matrix element
which still has large (non-perturbative) uncertainties.
A significant theoretical progress would be definitely welcome, as
$\varepsilon'/\varepsilon$ has been found to be potentially very constraining
for the LHT parameters and in particular for rare kaon decays.
With an accurate theoretical prediction at hand, $\varepsilon'/\varepsilon$
would become a strong constraint also for the MSSM, as pointed out in \cite{Buras:1999da,Buras:1998ed}.

We finally discuss the results found in \cite{Promberger:2007py} for rare kaon decays within the
3-3-1 ($Z'$) model. 
As discussed in Section \ref{subsec:2.3}, in this model significant NP effects
are expected in Flavour Physics as the heavy gauge boson $Z'$ can transmit
flavour changing neutral currents at tree-level.
We note that these tree-level effects are due to the peculiar role of the
third quark generation, set in a different $SU(3)_L$ representation w.r.t.
the first two.
They are, instead, absent in ew observables that turn out to 
induce only mild  constraints on the NP scale.
Significant NP effects are found in $\KLnu$ and $\Kp$ showing two possible
branches in the $Br(\Kp)$-$Br(\KLnu)$ plane, with a pattern similar to the LHT
one (see fig. \ref{fig:LHT}).
Here, however, smaller enhancements are allowed (up to a factor $4$ and $2$ for
$Br(\KLnu)$ and $Br(\Kp)$, respectively), because of the leptophobic nature of 
 the $Z' \bar \ell \ell$ coupling, suppressed by the weak mixing angle as
 $\sqrt{1-\sin^2 \theta_W}$.
Concerning $\KLl$, enhancements up to a factor $3$ w.r.t. the SM result to
be possible and strong correlations are found between the electron and muon
modes and with $\KLnu$.

\vspace*{0.2cm}
In conclusion, we  point out that while within MFV only small ($<30$
\%) NP effects can be expected in $\KLnu$, $\Kp$ and $\KLl$, much larger
(up to an order of magnitude) effects are allowed in non-MFV models, in
particular in the MSSM and in the LHT model.
Rare kaon decays clearly represent powerful tools to look
for NP and to investigate its flavour structure.

\end{document}